\documentstyle[epsf,aps]{revtex}
\begin{document}
\begin{flushright}
CU-TP-825 \\
DESY 97-072
\end{flushright}
\begin{center}
{\Large\bf Unitarity Corrections and High Field Strengths
in High Energy Hard Collisions}
\end{center}

\vskip 20pt

\begin{center}
Yuri V. Kovchegov$^{1,a}$ \footnotetext{a. This research is supported
in part by the Department of Energy under grant DE-FG02-94ER 40819.},
A.H. Mueller$^{1,a}$ and Samuel Wallon$^{2,b}$ \footnotetext{b.
Alexander von Humboldt Fellow.}
\end{center}

\begin{center}
$^1${\it Department of Physics, Columbia University \\ 
New York, New York 10027, USA}
\vskip 10pt 
{$^2$ II. {\it Institut f\"ur Theoretische Physik \\
Universit\"at Hamburg \\ Hamburg, Germany}}
\end{center}

\begin{abstract}
Unitarity corrections to the BFKL description of high energy hard
scattering are viewed in large $N_c$ QCD in light-cone quantization.
In a center of mass frame unitarity corrections to high energy hard
scattering are manifestly perturbatively calculable and unrelated to
questions of parton saturation. In a frame where one of the hadrons is
initially at rest unitarity corrections are related to parton
saturation effects and involve potential strengths $A_\mu \sim 1/g.$
In such a frame we describe the high energy scattering in terms of the
expectation value of a Wilson loop.  The large potentials $A_\mu \sim
1/g$ are shown to be pure gauge terms allowing perturbation theory to
again describe unitarity corrections and parton saturation effects.
Genuine nonperturbative effects only come in at energies well beyond
those energies where unitarity constraints first become important.
\end{abstract}

\newcommand{\stackeven}[2]{{{}_{\displaystyle{#1}}\atop\displaystyle{#2}}}
\newcommand{\lsim}{\stackeven{<}{\sim}}
\newcommand{\gsim}{\stackeven{>}{\sim}}

\section{\bf Introduction}

Our object in this paper is to describe a one (hard) scale high energy
scattering in QCD at those energies where unitarity corrections begin
to be important.  In particular, we focus on the question of the
perturbative calculability of such processes.  As our prototype
reaction we consider high energy heavy onium-heavy onium scattering
where the hard scale is given by the inverse onium radius, the only
relevant transverse momentum scale for the total cross section or for
a near forward elastic scattering.  We could equally well focus on
more directly physical processes such as the total virtual
photon-virtual photon cross section \cite{Ginzburg,Brodsky}
$\gamma^*(Q) + \gamma^*(Q) \to$ hadrons, or on associated jet
measurements either in electron-proton collisions
\cite{Tang,Bartels,Kwiecinski} or in proton-antiproton collisions
\cite{MN}, however unitarity issues are especially sharp in
onium-onium scattering since the onium is a bound state in QCD.

The energy dependence of an onium-onium collision is governed by the
BFKL equation \cite{Bali,Kuraev,Lipatov} , appropriate for single
scale high energy collisions.  If the onium is sufficiently small one
can neglect diffusion and running coupling effects and use the BFKL
equation up to those energies where unitarity corrections become
important.  We shall also use the large $N_c$ limit of QCD where one
can view the light-cone wavefunction of a high energy onium as a
collection of color dipoles \cite{Mueller,MP,Nikolaev}.  An
onium-onium collision, in the BFKL approximation, consists of a
scattering of a single dipole in each of the onia by means of one
gluon exchange.

The BFKL approximation gives an $S$-matrix for onium-onium elastic
scattering at impact parameter $b$\ and \ rapidity\ $Y$\ behaving as
\begin{eqnarray*}
S(Y,b) \sim \alpha^2 {e^{(\alpha_P-1)Y}\over [{7\over 2} \alpha
N_c\zeta(3)Y]^{3/2}}, \nonumber 
\end{eqnarray*}
a behavior which violates the unitarity bound $\vert S(Y,b) \vert \leq
1$ when Y becomes very large.  The picture of unitarity corrections is
very different in different Lorentz frames in light-cone quantized QCD
\cite{MS} in light-cone gauge.  In the center of mass frame unitarity
corrections become large when dipole densities, and corresponding
field strengths $A_\mu$, are relatively small. Unitarity bounds are
guaranteed by multiple dipole-dipole scattering \cite{Mueller,Salam}
in much the same way as occurs for nucleus- nucleus scattering in the
Glauber model.  One difference, as emphasized below, is that the
multiple dipole scattering must be done for a fixed wavefunction
configuration with a sum over the different configurations performed
as a last step.  The higher number of scatterings correspond to higher
orders in $1/N_c.$ In the center of mass system unitarity corrections
are quite straightforward and numerical simulations have already been
done \cite{Salam}.  Unitarity limits are enforced at rapidities well
below those where field strengths become large and parton saturation
effects are important.

It is instructive to view unitarity corrections in a different Lorentz
frame \cite{MS} where one of the onia, say the right-mover, carries
almost all the available rapidity with the left-moving onium having a
high enough rapidity to be relativistic but not so much that higher
(gluonic) left-moving Fock components must be considered in the
scattering.  In this frame, called the L-frame, the scattering is of a
right-moving onium with a fully developed wavefunction of many gluons
colliding with a left-moving onium whose wavefunction is simply a
heavy quark- antiquark pair.  The S-matrix can be written as an
average of Wilson loops in the right-moving state as indicated in
(30). From this representation it becomes clear that unitarity
corrections become important in the L-frame when $A_\mu\sim 1/g.$ In
this frame unitarity is associated with high potential strengths and
significant parton saturation.  However,it is suggested that these
large values of $A_\mu$ do not yet signal nonperturbative QCD effects
since the large terms in $A_\mu$ are purely gauge terms.  In order to
see genuine nonperturbative QCD effects one must go to rapidities well
beyond (about twice as large) those rapidities where unitarity
corrections first become important.

\section{\bf Unitarity Corrections in the Center of Mass Frame}

In this section, we review high energy heavy onium -- heavy onium
\cite{Mueller,MP,Nikolaev,MS,Salam} scattering in large $N_c$ QCD
beginning at moderate energies where the two gluon exchange
approximation is valid, through the energy range where the BFKL
equation governs the scattering, up to energies where unitarity
corrections to the BFKL equation become important,into the energy
domain where unitarity corrections are very strong and the forward
S-matrix elements become nearly zero (blackness), and ending at
energies so high that the whole perturbative picture breaks down and
high field strengths (parton saturation) occur.

Suppose $\Phi(x_{01},z)$ is the square of the heavy quark light-cone
wavefunction of the onium with $x_{01} = \vert \b{x}_1-\b{x}_0\vert$
the transverse coordinate separation between the heavy quark, at
$\b{x}_0,$ and the heavy antiquark, at $\b{x}_1$ and with \ z\ the
fraction of the onium's longitudinal momentum carried by the quark.
Then in the two gluon exchange approximation the forward elastic high
energy scattering amplitude is

$$A = {i\over 2} \int d^2 x_{01} d^2 x_{01}^\prime \int_0^1 dz
dz^\prime \Phi (x_{01}, z) \Phi (x_{01}^\prime, z^\prime) \sigma_{DD}
(x_{01}, x_{01}^\prime)\eqno(1)$$

\noindent where the dipole-dipole cross section, $\sigma_{DD}$, is

$$\sigma_{DD}(x_{01}, x_{01}^\prime) = 2\pi \alpha^2 x_<^2 [1+\ln 
(x_>/x_<)]\eqno(2)$$

\noindent with $x_<$ the smaller of $x_{01}$ and $x_{01}^\prime$ and
with $x_>$ the larger of these quantities.  The onium-onium cross
section is

$$\sigma = 2 \, \mbox{Im} A.\eqno(3)$$

\noindent which is independent of energy at high energies.  The
$\alpha$ in (2) is $\alpha(R)$ where

$$R = \frac{1}{2}\int d^2 x_{01} \int_0^1 dz x_{01} \Phi (x_{01},
z)\eqno(4)$$

\noindent is the average light-cone radius of the heavy quark part of
the onium wavefunction.  If $R \, \Lambda_{QCD}$ is sufficiently
small, running coupling effects will not be important in onium-onium
scattering up to energies where saturation sets in and the whole
perturbative picture of high energy scattering breaks down.

If $Y = \ln (s/M^2)$ is the relative rapidity of the two colliding
onia then the two gluon approximation breaks down when $\alpha Y$ is
of order 1.  In this regime the BFKL equation, an equation which sums
all $(\alpha Y)^n$ terms, governs high energy onium-onium scattering.
In the dipole picture of high energy scattering one writes the
scattering amplitude at impact parameter b as \cite{Mueller}

$$A(b,Y) = - i \int d^2 x_{01} d^2 x_{01}^\prime \int_0^1 dz dz^\prime
\Phi (x_{01}, z) \Phi (x'_{01},z^\prime) F(x_{01}, x_{01}^\prime,
b,Y)\eqno(5)$$

\noindent where in the BFKL approximation $F = F^{(1)}$ and

$$F^{(1)} = - {1\over 2} \int_0^\infty {dx\over x}\ \ {dx^\prime\over
x^\prime}\ d^2b_1 n(x_{01}, x, b_1, Y/2) n(x_{01}^\prime,
x^\prime,\vert \b{b}-\b{b}_1\vert,Y/2)
\sigma_{DD}(x,x^\prime).\eqno(6)$$

\noindent $n(x_{01}, x, b_1, Y/2)$ is the number density of dipoles of
size\ x\ whose center is a distance $b_1$ from the center of the heavy
quark-heavy antiquark system, $x_{01}.$ The forward elastic amplitude
is

$$A(Y) = \int d^2b A(b,Y).\eqno(7)$$

For $x/b, x_{01}/b \ll 1$ \cite{Mueller,Salam,NW}

$$n(x_{01}, x,b,Y) = {x_{01}\over 4 x b^2}\ \ln \left({16 b^2\over
x_{01}x}\right) \ {\exp\{(\alpha_P-1)Y-a(Y)/2\ \ln^2(16
b^2/x_{01}x)\}\over [{7\over 2} \alpha N_c \zeta(3)Y]^{3/2}}\eqno(8)$$

\noindent with $\alpha_P-1=(4\alpha N_c/\pi) \ln 2$ and $a^{-1} (Y) =
\frac{7 \alpha N_c \zeta(3) Y}{\pi}$.  Using (2) and (8) in (6) gives

$$F^{(1)} = - \pi\alpha^2 {x_{01}x_{01}^\prime\over b^2}\ \ln
\left({16 b^2\over x_{01} x_{01}^\prime}\right)
{exp\{(\alpha_P-1)Y-{a\over 2}\ln^2({16 b^2\over
x_{01}x_{01}^\prime})\}\over [{7\over 2} \alpha
N_c\zeta(3)Y]^{3/2}}\eqno(9)$$

\noindent when $x_{01}/b, x'_{01}/b \ll 1.$ The resulting onium-onium
cross section

$$\sigma = 16\pi R^2\alpha^2 {e^{(\alpha_P-1)Y}\over \sqrt{{7\over
2}\alpha N_c\zeta(3)Y}}\eqno(10)$$

\noindent comes from impact parameters satisfying $\ln \left(
{b^2\over R^2}\right) \sim {\sqrt{14\alpha N_c\zeta(3)Y/\pi}}.$ From
(3), (5), (7) and (9) one obtains

$${d\sigma\over d^2b} = 8\pi R^2 \alpha^2 \ln \left( 16 b^2/R^2
\right) {\exp\{(\alpha_P-1) Y- {a\over 2} \ln^2({16 b^2\over
R^2})\}\over [{7\over 2} \alpha N_c\zeta(3)Y]^{3/2}b^2}\eqno(11)$$

\noindent so long as $R^2/b^2 \ll 1$.  Since

$${d\sigma\over d^2b} = 1 - \mbox{Re} \ S(b), \eqno(12)$$

\noindent with $S(b)$ the $S$-matrix for onium-onium scattering at
impact parameter b, (10) and (11) violate the unitarity bound
${d\sigma\over d^2b} \leq 2$ when $\alpha Y$ becomes large.

The dipole picture of BFKL behavior is very simple.  At high energies
each of the onia consists of a heavy quark-antiquark pair along with a
(generally) large number of soft gluons ordered in longitudinal
momentum.  In the large $N_c$ limit each gluon can be viewed as a
quark- antiquark pair and the onium wavefunction then can be viewed as
a collection of color dipoles.  In the scattering of two onia, in the
BFKL approximation, a single right-moving dipole from the right-moving
onium scatters off a single left-moving dipole in the left-moving
onium, with the two gluon exchange cross section given in (2), leading
to the parton-like expression (6).

Roughly speaking, one expects unitarity corrections to the BFKL
approximation to become important when ${d\sigma\over d^2b} \approx
1.$ From (2) and (6) this happens at values of Y when $n \sim
1/\alpha$ and one might expect that the leading logarithmic
approximation to the onium wavefunction would start to break down.
However, it is straightforward to see that when ${d\sigma\over d^2 b}=
1$

$$x^2 n \sim {x\over \alpha b}.\eqno(13)$$

\noindent Since $x/b \ll 1$ for those dipole sizes dominating the
integral in (6) the onia wavefunctions are still quite dilute when
unitarity corrections begin to become important.  The situation is
somewhat analogous to the scattering of two large, but dilute,
"nuclei."  If the nuclei have atomic number A and radius R then, in
the single scattering approximation

$$\sigma_{AA} = A^2\sigma_0 = 4\pi R^2 \rho^2\left({4\pi\over 9}\
{R^4\over \sigma_0^2}\right), \eqno(14)$$

\noindent where $\sigma_0$ is the nucleon-nucleon cross section and
$\rho$ is the packing fraction defined by

$$\rho = {A\sigma_0^{3/2}\over (4/3) \pi R^3}.\eqno(15)$$

From (14) one sees that the unitarity limit, $4\pi R^2,$ is reached
for $\sigma_{AA}$ for very small packing fraction if $R^2/\sigma_0 \gg
1.$ In this nuclear example it is clear how to impose unitarity.
Since the nucleus is dilute one need only do multiple scattering
quantum mechanically to obtain a correct forward scattering amplitude.

The unitarity limit is reached for onium-onium scattering in the
center of mass frame also when the dipole densities in the onia are
small so that the two dipoles involved in the scattering can be viewed
as not interacting with any other dipoles. It would seem that
unitarity can be imposed simply by including two and more scatterings
of left-moving dipoles with a similar number of right-moving dipoles.
And this is indeed the case \cite{Mueller,Salam}.  There are, however,
two differences from the ``nuclear'' example.  (i) The dipoles have
various sizes whereas the nucleons in our nuclear example all have the
same size.  This is a technical problem which makes an analytical
discussion of the problem difficult but does not present any problems
of principle \cite{Mueller}.  (ii) The wavefunction of an onium has
large fluctuations in the number of dipoles present.  Thus the single
scattering term in the S-matrix is dominated by very different parts
of the wavefunction than is the double scattering term which in turn
samples a very different part of the wavefunction than the triple
scattering term, etc. This causes the normal multiple scattering
series (multiple pomeron exchange) to diverge factorially
\cite{Mueller,Salam}.  This means that one must sum the multiple
scatterings of the dipoles in the colliding onia for a fixed
wavefunction configuration and only after this sum has been done can
the average over the different configurations of the onia
wavefunctions be done.  Schematically,

$$S(Y,b) = \sum_\phi e^{-f(\phi)}P_\phi(Y)\eqno(16)$$

\noindent where $\phi$ labels configuration of dipoles in each of the
colliding onia.  That is $\phi$ labels the number of dipoles in each
onium along with the size of each of the dipoles and the impact
parameter of the center of each of the dipoles with respect to the
center of the heavy quark-antiquark pair.  $P_\phi$ gives the
probability of configuration $\phi$ while $f(\phi)$ is proportional to
the probability of a dipole-dipole scattering in the configuration
$\phi.$ b is the impact parameter of the onium-onium collision. If one
expands the exponential in (16) and interchanges the sum with $\phi$

$$S(Y,b) = \sum_{n=1}^\infty (-1)^n S_n(Y,b)\eqno(17)$$

\noindent with

$$S_n=\sum_\phi {1\over n!} [f(\phi)]^n P_\phi(Y)\eqno(18)$$

\noindent gives the usual multiple scattering series.  However, simple
models \cite{Mueller} and numerical calculations in QCD \cite{Salam}
show that $S_n\sim n!$ so that the multiple scattering series does not
converge.  The problem is that rare configurations which have a large
value of $f(\phi)$ contribute very much to $S_n,$ for large $n,$ but
very little to $S(Y,b)$ as given in (16).

Numerical simulations have been carried out within the dipole
framework of high energy onium-onium scattering which follow the BFKL
behavior given in (10) and (11) into the regime of $Y$-values where
unitarity corrections due to multiple scattering are important.  These
simulations suggest that \cite{MS}

$$S(Y,b) \sim e^{-c(Y-Y(b))^2}\eqno(19)$$

\noindent for \ $Y$\ much greater than $Y(b),$ the value of $Y$ at
which unitarity corrections become important. The behavior given in
(19) is a much slower decrease of $S$ with $Y$ than would be expected
from an eikonal picture and corresponds to the $S$ matrix being
dominated by rare configurations consisting of many fewer dipoles than
the average at the $Y$ value being considered.  The behavior (19) is
also expected on theoretical grounds.

If $Y_0$ is the rapidity at which unitarity corrections begin to be
important for $b=0$ collisions then at $Y \approx 2Y_0$ one expects
perturbation theory to break down completely and a new regime of high
field strength QCD to emerge.  This can be seen from the following
rough argument: In (6) neglect the integrations over $dx dx^\prime$
and set $d^2b_1=R^2$ for a $b=0$ collision.  (The neglected
integrations only give logarithmic prefactors in any case.)  From (6)
it is clear that unitarity corrections become strong when $R^2n(Y_0/2)
\sim 1/\alpha.$ However, at this value of $Y_0$ the onium wavefunction
is still relatively dilute.  Indeed, if one scatters a single low
momentum left-moving dipole on a fully developed right- moving onium
state of rapidity $Y_0/2$ the cross section is of order $\alpha.$ The
interaction of the fully developed right-moving onium state with a
slightly left-moving dipole is the same as the interaction with a
slightly right-moving dipole which could be part of the onium
wavefunction.  It is only when $R^2n \approx 1/\alpha^2$ that a given
low momentum dipole in an onium wavefunction begins to strongly
interact with some of the dipoles of higher rapidity.  Thus in a
center of mass collision at $Y=2Y_0$ all the soft dipoles in both the
right-moving and left-moving onia suffer a strong interaction and a
good portion of these dipoles will appear as freed gluons in the
collision leading to a high field strength initial state,
$R^2F_{\mu\nu} \sim 1/g,$ of the collision.  However, at such high
field strengths our control over the collision is lost because
perturbation theory breaks down.  (This regime is usually referred to
as the saturation regime because a limit to the growth of parton
densities is expected to occur.)  Whether there is some classical, or
semiclassical, solution which governs such collisions is an
interesting and challenging problem which, however,goes far beyond the
purposes of the present work.

\section{Scattering in an Almost Rest System}
In this section we shall view the forward scattering of two onia in a
frame where one of the onia, say the right-moving onium, has almost
all the available rapidity while the left- moving onium has only
enough rapidity to make it move relativistically, but not enough so
that one is required to add gluons to its wavefunction \cite{MS}.
Thus the left-moving onium in this almost rest system, which we shall
call the L-frame, consists only of a heavy quark and a heavy
antiquark, a single dipole.

\subsection{The BFKL Approximation in the L-Frame}

The BFKL approximation is straightforward to implement in the L-frame.
In (6) one just makes the replacement

$$n(x_{01}^\prime, x^\prime,\vert \b{b}-\b{b}_1 \vert, Y/2) \to
n(x_{01}^\prime, x^\prime, \vert \b{b}-\b{b}_1 \vert,Y)\eqno(20)$$

$$n(x_{01}, x, \b{b}_1, Y/2) \to x\delta(x-x_{01})
\delta(\b{b}_1).\eqno(21)$$

\noindent Eq.(21) simply says that the left-moving onium consists of a
single dipole, the heavy quark-heavy antiquark pair.  The result (9)
emerges as is easy to check.

However, it is useful to view the process in a somewhat different way
\cite{McLerran,Jamal,Yuri}.  Let us view the interaction of the two
onia as they pass each other as due to the interaction of the left-
moving heavy quark-antiquark pair with the color field of the
right-moving collection of color dipoles making up the right-moving
onium.  Thus as the onia pass each other we consider a fixed
configuration of dipoles in the right-moving onia.  Each dipole gives
a classical color field (We take traces over the colors in the
individual dipoles later.), and because of the large $N_c$
approximation the color fields coming from the different dipoles do
not interfere. The interaction with the left-moving heavy
quark-antiquark pair with a dipole in the right-moving onium of size
$x^\prime$ at impact parameter $b_1$ from the center of the
left-moving heavy quark- antiquark pair is

$$W_2(\b{x}^\prime, \b{x}_{01}, \b{b}_1) = {1\over N_c}\left[tr P \exp
\left( ig \oint A\cdot dx \right) \right]_2\eqno(22)$$

\noindent where the integration goes around a closed Wilson loop in
the $x_\perp, x_-$ plane as illustrated in Fig.1. Indeed, explicit
evaluation identifies $W_2$ with $- \frac{1}{2 \pi} f(\underline{b}_1,
\underline{x}', \underline{x}_{01})$ of Ref. \cite{Salam}. The $A_\mu$
in (22) is the field due to the color dipole of size $x^\prime$ and
the subscript 2 in (22) means that we are only taking the order $g^2$
term.  The exact values of $x_ -^\ell$ and $x_-^r$ for the Wilson loop
do not matter so long as $x _-^r-x_- ^\ell \gsim R$ and $x_-^r >0,
x_-^\ell <0$ with neither $x_-^r$ or $x_-^\ell$ being too close to
zero.  (We suppose the centers of the two onia pass each other at $x_+
= x_- = 0.)$ The field, not yet counting color factors, due to the
dipole $x^\prime$, consisting of a ``quark'' at ${\underline x}'_a$
and an ``antiquark'' at ${\underline x}'_b$, is \cite{Yuri}

$$A_\mu(x) = g{\delta(x_-)\over 4\pi} \ln \left( {({\underline
x}-{\underline x}_b^\prime)^2\over
(\underline{x}-\underline{x}_a^\prime)^2} \right) {\overline \eta}_\mu
\eqno(23)$$

in covariant gauge and

$$A_\mu(x) = {g \over {4\pi}} \epsilon (x_-) \left(
{(x-x_a^\prime)_\mu^\perp \over {(\b{x}-\b{x}_a^\prime)^2}} -
{(x-x_b^\prime)_\mu^\perp \over {(\b{x}-\b{x}_b^\prime)^2}} \right)
\eqno(24)$$

\noindent in principal-value light-cone gauge.  In (23) $\overline
\eta$ is such that $\overline \eta \cdot v=v_-$ for any four-vector
$v_\mu$ while $x_\mu^\perp$ in (24) means that only the $\mu = 1,2$
components are nonzero.  From the forms of $A_\mu$ in (23) and (24) it
is clear that the exact values of $x_-^r$ and $x_-^\ell$ will not
matter when evaluating the Wilson loop in (22).

When $x_{01}/b, x_{01}^\prime/b \ll 1$ it is straightforward to check
that

$$F^{(1)}(x_{01}, x_{01}^\prime , b,Y) = \int {d^2x^\prime\over 2\pi
x^{^\prime 2}} d^2 b_1 n(x_{01}^\prime,x',\vert\b{b}-\b{b}_1 \vert,Y)
W_2(\b{x}^\prime,\b{x}_{01},\b{b}_1)\eqno(25)$$

\noindent agrees with (9).  Eq.(25) is in fact true for all \ b\ in
the leading logarithmic (BFKL) approximation. (Also, the lowest order
approximation, (21), for $n$ allows one to identify $W_2$ as the
dipole-dipole scattering amplitude at a definite impact parameter.)
One can use (25) to find out what are the typical values of the fields
$A_\mu$ in the right-moving onium.  From (22) and (25)

$${1\over N_c} tr \left< \left( g\oint A \cdot dx \right)^2 \right> =
- 2\int d^2 x_{01}^\prime \int_0^1 dz^\prime \Phi(x_{01}^\prime,
z^\prime) F^{(1)}(x_{01}^\prime, x_{01}, b,Y)\eqno(26)$$

\noindent where the left-hand side of (26) is the expectation in the
right-moving onium state of the square of the loop integral.  This is
now the field due to all of the dipoles in the right-moving onium.
In the BFKL approximation (26) is a gauge invariant equation.  Using (9)
one finds

$${1\over N_c} tr \left< \left( g\oint A\cdot dx \right)^2 \right> =
4\pi \alpha^2{ R x_{01}\over b^2} \ln \left({16 b^2\over R
x_{01}}\right) {\exp\{(\alpha_P-1)Y - {a \over 2} \ln^2({16 b^2\over R
x_{01}})\}\over [{7\over 2} \alpha N_c\zeta(3)Y]^{3/2}}.\eqno(27)$$

In light-cone gauge the integration of the $A_\mu$ field goes over
distances $x_{01}$ so that the integration picks out those parts of
$A_\mu$ having wavelength $x_{01}$ or greater.  Define the left-hand
side of (27) to be ${4\over N_c} x_{01}^2 g^2 \sum_i \left<
(A_\perp^i)^2 \right>.$ (The factor of 4 is included because,
according to (24) there are equal contributions coming from $x_- < 0$
and from $x_- > 0$ in the integral on the left-hand side of (27).)
Then,

$$\left< (A_\perp^i)^2 \right> = {\alpha R \over 2 N_c x_{01} b^2} \ln
\left({16 b^2 \over R x_{01}}\right) {\exp \left[ {(\alpha_P-1)Y - {a
\over 2} \ln^2({16 b^2\over R x_{01}})} \right] \over [{7 \over 2}
\alpha N_c \zeta(3)Y]^{3/2}} \eqno(28)$$

\noindent gives the average value for the square of a fixed color
component and a fixed transverse spatial component of $A_\mu$ in
light-cone gauge when the field is measured a distance b from the
center of the fast moving onium and if the measurement averages over
transverse distances of size $x_{01}.$ When $\alpha Y$ is of order\ 1\
the field $A_\mu$ is of size g.  Taking $x_{01} = 2R$ and using (11)

$$4R^2 \left< (A_\perp^i)^2 \right> = {1\over 8\pi \alpha N_c}\ {d
\sigma\over d^2b}\eqno(29)$$

\noindent so that the unitarity limit, in the BFKL approximation, is
reached when typical field strengths reach $A \sim (g{\sqrt{N
_c}})^{-1}.$ Hence, when light-cone field strengths reach values of
the order of $1/g$ one expects strong unitarity corrections to
occur. Thus, in the L-frame the onset of unitarity corrections to the
BFKL approximation occurs at exactly the same rapidity that field
strengths reach size $1/g.$ This is in contrast to the center of mass
frame where unitarity corrections become important when field
strengths are of size $g^0.$

\subsection{Unitarity Corrections in the L-Frame}

When field strengths reach a size $1/g$ the complications of doing
calculations in the L- Frame are of two varieties.  First, one can no
longer expect the two gluon approximation to be valid for connecting
the right-moving system of a heavy quark-antiquark pair and its
accompanying right-moving gluons to the left-moving heavy
quark-antiquark pair.  Secondly, one can no longer expect the
right-moving system to consist only of dipoles.  When $A_\mu^i \sim
(g{\sqrt{N_c}})^{-1}$ the large $N_c$ expansion begins to break down
and one can expect more complicated color structures, quadrupoles,
sextupoles, etc. to become important in addition to the color dipoles
corresponding to the BFKL approximation.  In the center of mass system
the breakdown of the large $N_c$ expansion shows up in the multiple
scattering in (16) and, more explicitly, in (17) and (18).

We continue to find it convenient to view the scattering as the
interaction of the left- moving heavy quark-antiquark pair (dipole)
with the color field of the right-moving system.  Thus from (5), with
$S(Y,b) = 1 + i A(Y,b)$

$$S(Y,b) = {1\over N_c} \int d^2 x_{01} \int_0^1 dz_1 \Phi(x_{01},
z_1) tr \left< P e^{ig\oint A_\mu dx_\mu} \right>_b\eqno(30)$$

\noindent where the contour of integration for $dx_\mu$ is the same as
before, and shown in Fig.1, while the average in (30) means that
$A_\mu$ is the field in the high momentum right-moving onium at fixed
impact parameter \ b\ from the center of the right-moving heavy quark-
antiquark pair and with averages being done over all the numbers and
distributions of gluons (dipoles, quadrupoles, etc.) of the
right-moving system making up the wavefunction of the onium.  That the
Wilson loop correctly expresses the scattering amplitude is apparent
in covariant gauge where the interaction with the left-moving
quark-antiquark pair is causal and occurs with the horizontal lines
shown in Fig. 1.  Since the left-moving heavy quark-antiquark pair is
frozen, in transverse coordinates, during the passage through the
field due to the right-moving system, the integral in (30) just
represents the eikonal interaction of that field with the heavy quark
at $\b{x}_0$ and with the heavy antiquark at $\b{x}_1$.

In the BFKL approximation the fields from the right-moving system come
from independent dipoles and (30) is replaced by (25).  In general
(30) shows that when one reaches rapidities where $A_\mu^i \sim
(g{\sqrt{N_c}})^{-1}$, and unitarity corrections become strong, what
is being determined in the scattering is the expectation of a Wilson
loop.  The expectations of the various sized Wilson loops contain the
essential information of QCD.  Very high energy scattering allows one
to test QCD in a regime where $g$ is small, corresponding to the small
onium size, but where $g A$ is not small.

Suppose we take $Y=Y_0,$ the rapidity at which unitarity corrections
begin to be important for a zero impact parameter onium-onium
collision.  At such rapidities $g A_\mu$ is of order 1 in (30), and
one might expect that essential nonperturbative physics would be
necessary to evaluate A.  However, at a corresponding rapidity we
found that perturbation theory works well in the center of mass system
where $g A_\mu$ is of order $g$. How does one reconcile this?  The
resolution is that the field $A_\mu \sim 1/g$ in (30) when unitarity
corrections become important, but that the large part of this field is
purely gauge.  The potential $A_\mu$ comes essentially from a
superposition of independent fields of the type shown in (24).  The
potential given in (24) leads to a field strength

$$F_{+i} = {g\over 2\pi} \delta(x_-) \left[{(x-x_a^\prime)_i\over
(\b{x}- \b{x}_a^\prime)^2} \ -\ {(x-x_b^\prime)_i \over
(\b{x}-\b{x}_b^\prime)^2}\right].\eqno(31)$$

\noindent In reality the $\delta(x_-)$ should be smeared out
\cite{Jamal,Yuri} over an extent $\Delta x_-\propto R e^{-y_{ab}}$
where $y_{ab}$ is the smallest of the rapidities of the gluons making
up the ``quark a'' and the ``antiquark b'' which is the source of (24)
and (31).  Thus, although the potential $A_\mu$ is of size $1/g$ when
$Y = Y_0$ the $F_{+i}$'s which come from the different dipoles which
serve as sources for \ A\ have differing longitudinal extent $\Delta
x_-$ so that there is no fixed coherence distance $\Delta x_-$ over
which $F_{+i}$ is of size $1/g.$ This means that the $1/g$ part of
$A_\mu$ is purely a gauge.  (The largest number of sources having a
fixed coherence length are those dipoles which are (partly) made up of
gluons having rapidity on the order of 1.  When $Y = Y_0$ there is on
the order of $N_0 \sim 1/\alpha$ such sources.  These sources lead to
a non-gauge part of $A_\mu$ of size $A_\mu \sim {\sqrt{N_0}}\cdot g
\sim 0(1)$ which is the maximum non-gauge field present at $Y=Y_0.$)
Nevertheless, the determination of the $A_\mu$ to use in (30) does
involve nonlinear (gauge) interactions because $gA$ is not small. It
is not a trivial problem to evaluate $A_\mu$ for an onium state having
rapidity $Y \gsim Y_0.$ For the somewhat simpler problem of a large
nucleus where $A \sim 1/g$ comes from different (non-overlapping)
nucleons, the corresponding pure gauge field, which could be used in
(30) to give hadron nucleus scattering, has been calculated in
Refs. \cite{Jamal} and \cite{Yuri}.

Thus at those rapidities where unitarity corrections begin to be
important the potential $A_\mu$ becomes of size $1/g$ but this large
part of the potential is purely gauge.  As one further increases $Y$
the rate of growth of $A_\mu$ will be considerably slower than for $Y
< Y_0$ in keeping with (30) leading to (19).  This slowing down of
$A_\mu$ corresponds to ``parton saturation'' and in principle is
perturbatively calculable.  In the L-frame parton saturation is
intimately related to the rate at which unitarity (blackness) sets in
as $Y$ becomes greater than $Y_0$ \cite{MS}. However, as we have seen
in Sec.2 one can calculate unitarity corrections in the center of mass
system where field strengths remain weak and saturation effects are
not important.

Finally, we have argued that at values of $Y$ on the order of $Y_0,$
where unitarity corrections first become important, the wavefunction
of a heavy onium is still a perturbative object even though $A_\mu
\sim 1/g.$ Thus the general probe described by the Wilson loop in (30)
and corresponding to scattering by an undeveloped onium state does not
require knowledge of nonperturbative QCD.

However, suppose we scatter, at zero impact parameter, a right-moving
onium, having $Y \approx Y_0,$ on a left-moving onium, having $Y
\approx Y_0.$ In the central unit of rapidity there are on the order
of $1/\alpha$ gluons, corresponding to $A \sim 1/g,$ which will be
freed during the collision.  These gluons lead to field strengths
$F_{\mu\nu} \sim 1/g$ and now genuine nonperturbative QCD effects can
be expected.  Thus the genuine nonperturbative effects appear to be
more connected with production in the collision process than with the
wavefunction of a high energy hadron, or at least they show up at
lower rapidities in the collision than in wavefunctions entering the
collision.

\begin{figure}
\begin{center}
\epsfxsize=10cm
\epsfysize=4cm
\leavevmode
\hbox{ \epsffile{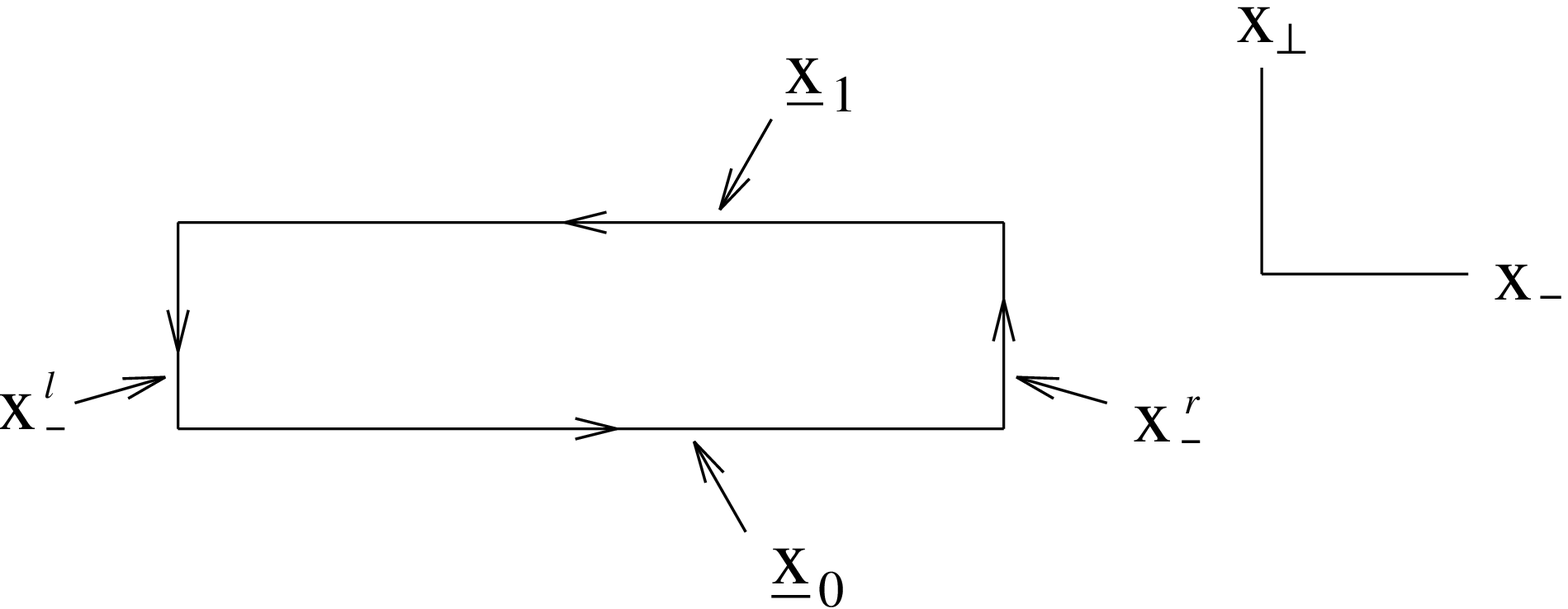}}
\end{center}
\caption{}
\label{}
\end{figure}

\end{document}